# DESIGN GUIDELINE FOR MINIMIZING SPACE-CHARGE-INDUCED EMITTANCE GROWTH

Chuan Zhang [#], GSI Helmholtz Center for Heavy Ion Research, Planckstr. 1, Darmstadt, Germany

*Abstract*

Space-charge-induced emittance growth is a big concern for designing low energy and high intensity linacs. The Equipartitioning Principle was introduced to minimize space-charge-induced emittance growth by removing free energy between the transverse and longitudinal degrees of freedom. In this study, a different design guideline is being proposed. It suggests to hold the ratio of longitudinal emittance to transverse emittance around one and take advantage of low emittance transfer for minimizing emittance growth. Using a high intensity RFQ accelerator as an example, a comparison between the two design methods has been made.

## BACKGROUND

In high intensity linacs, space-charge-induced emittance growth is a big concern, especially at low energy.

As early as in 1950s, I. M. Kapchinsky and V. V. Vladimirsky studied the effect of space charge on the transverse motion using a uniformly charged, infinitely long, elliptical-cylinder-like beam [1].

However, the longitudinal-transverse coupling was firstly identified by R. Chasman in 1968 as an important mechanism for space-charge-induced emittance growth in high current proton linacs, after the 6D numerical computations had been performed [2].

At the same time, P. M. Lapostolle proposed that the coupling-caused emittance growth could be minimized by equipartitioning [3].

In 1981, I. Hofmann reported the stability thresholds for different coupling modes calculated using the Vlasov equation for an initial Kapchinsky-Vladimirsky distribution with arbitrary emittance ratios, tune ratios, and intensity [4]. These thresholds originally obtained from continuous beams in an x-y geometry were immediately applied to the r-z geometry for understanding the longitudinal-transverse emittance transfer in bunched beams [5]. I. Hofmann visualized the thresholds in the form of charts and suggested that these stability charts can give a useful orientation for studying the longitudinal-transverse coupling in linacs [6].

Also in 1981, R. A. Jameson published the Equipartitioning Principle (EP) as well as the EP equation i.e. Eq. (1) and proposed minimizing space-charge-induced emittance growth by removing free energy between the transverse and longitudinal degrees of freedom [7]. He pioneered the application of Hofmann Stability Charts as important tools for designing linacs with minimum emittance growth.

___________________
# c.zhang@gsi.de

$$\frac{\varepsilon_l \sigma_l}{\varepsilon_t \sigma_t} = 1 \quad (1)$$

As an example, Fig. 1 shows one Hofmann chart with the tune ratio $\frac{\sigma_l}{\sigma_t}$ and the tune depression $\frac{\sigma}{\sigma_0}$ as the abscissa and the ordinate, respectively. The darker the purple color is, the higher the growth rate of exchange (also called as resonance) is. The main resonance peaks appear at the positions where $\frac{\sigma_l}{\sigma_t} = \frac{m}{n}$ (*m* and *n* are integers), e.g. $\frac{\sigma_l}{\sigma_t} = \frac{1}{2}$, $\frac{1}{1}$, $\frac{2}{1}$, etc.

On a Hofmann chart, the maximum spread of the safe tune depression is always available at a location where the EP equation is satisfied and the $\frac{\sigma_l}{\sigma_t} = \frac{\varepsilon_t}{\varepsilon_l}$ resonance peak expected to be present is "killed" (see the blue dashed line in Fig. 1). However, any deviation from this EP condition will result in the reappearance of the "killed" peak. The larger the deviation is, the more the peak regrows.

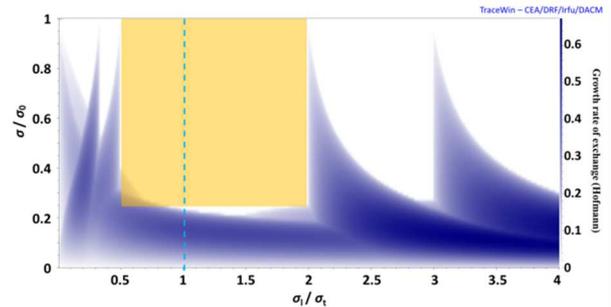

Figure 1: Hofmann chart for $\frac{\varepsilon_l}{\varepsilon_t} = 1.0$.

## NEW DESIGN GUIDLINE

In reality, emittance transfer can't be avoided completely. Therefore, it makes more sense to choose the clean area on the Hofmann charts instead of sticking on the EP lines for the beam motion.

In a previous study [8], it was found that the Hofmann chart with $\frac{\varepsilon_l}{\varepsilon_t} = 1.0$ can provide a quasi-rectangular clean area with very wide ranges of tune ratio ($\frac{\sigma_l}{\sigma_t} = 0.5 - 2.0$) and tune depression ($\frac{\sigma}{\sigma_0} = \sim 0.25 - 1.0$), respectively. To minimize the emittance transfer, it was recommended using this "safe rectangle" (see the orange marked area in Fig. 1) to the greatest extent for the beam motion [8]. For this purpose, one should try to hold the emittance ratio $\frac{\varepsilon_l}{\varepsilon_t}$ at one [8]. However, it is actually very difficult to be done perfectly in real machines due to different reasons e.g. errors.

Different than the proposal by the previous study, a new design guideline is being suggested to hold $\frac{\varepsilon_l}{\varepsilon_t}$ around one, which is more realistic. It can be also seen in [8] that the $\frac{\sigma_l}{\sigma_t} = 1.0$ resonance peak will regrow when a deviation from $\frac{\varepsilon_l}{\varepsilon_t} = 1.0$ starts. But fortunately, in the range of $\frac{\varepsilon_l}{\varepsilon_t} = 0.9 - 1.4$, the $\frac{\sigma_l}{\sigma_t} = 1.0$ peak is not significant and its growth rates are low. In addition, in this $\frac{\varepsilon_l}{\varepsilon_t}$ range, the growth rates for the $\frac{\sigma_l}{\sigma_t} \leq 0.5$ resonance peaks are decreasing with an increasing $\frac{\varepsilon_l}{\varepsilon_t}$ [8].

Besides allowing the low emittance transfer in the range $\frac{\varepsilon_l}{\varepsilon_t} = 0.9 - 1.4$, another new idea is to take advantage of them for minimizing emittance growth.

Typically, an RFQ input beam has very small energy spread $\Delta W_{in}$ but very large phase spread $\Delta \varphi_{in}$, so for RFQ beam dynamics design studies it is usually assumed that $\Delta W_{in} = 0$ and $\Delta \varphi_{in} = \pm 180°$, respectively, which leads to $\varepsilon_{l,in} = 0$. In the pre-bunching, $\varepsilon_l$ is being increased without significant emittance transfer, as the longitudinal and transverse beam sizes are still not comparable before the beam bunch is really formed. In the subsequent main bunching, the beam is further compressed longitudinally but the beam velocity is still low, so emittance transfer can occur from the longitudinal plane to the transverse ones. When the real acceleration starts, the transverse defocusing effect will be weakened naturally and emittance transfer can change the direction namely that emittance transfer occurs from the transverse planes to the longitudinal one. The part around the end of the main bunching is most critical for space charge, especially at high intensities.

Therefore, the strategy to use low emittance transfer for minimizing emittance growth is as follows:
- To choose a relatively large $\frac{\varepsilon_l}{\varepsilon_t}$ in the range $0.9 - 1.4$, say 1.3, for the end of the pre-bunching. Then the emittance transfer started with the main bunching will decrease $\frac{\varepsilon_l}{\varepsilon_t}$ to a value around 1.0.
- The beam motion in the most critical part for space charge should be kept inside the "safe rectangle".
- Afterwards, the emittance transfer will change the direction and increase $\frac{\varepsilon_l}{\varepsilon_t}$, but the $\frac{\sigma_l}{\sigma_t} \leq 0.5$ resonance peaks with low growth rates will not lead to a very significant emittance transfer.

## DESIGN AND SIMULATION

To apply the new design guideline, a 324 MHz, 3 MeV, 60 mA proton RFQ has been taken as an example. Table 1 lists its basic parameters. They are the same as those of the J-PARC epRFQ [9] which was designed as a "fully equipartitioned" machine. This allows a comparison between the two methods. Although the J-PARC epRFQ works with H$^-$ ions, the difference between H$^+$ and H$^-$ ions can be ignored from beam dynamics point of view. In addition, different than the changing inter-vane voltage $U$ adopted by the J-PARC epRFQ, a constant $U = 75$ kV has been chosen for the proton RFQ.

Table 1: Basic design parameters of the proton RFQ

| Parameter | Value |
|---|---|
| Frequency [MHz] | 324 |
| Input energy [keV] | 50 |
| Output energy [MeV] | 3 |
| Beam current [mA] | 60 |
| Input emittance $\varepsilon_{t, in, n., rms}$ [$\pi$ mm mrad] | 0.2 |

Following the new guideline, the beam dynamics design of the proton RFQ has been made by means of the New Four Section Procedure [10]. Fig. 2 shows the evolution of the main design parameters along the proton RFQ, where $a$ is the minimum electrode aperture, $m$ is the electrode modulation, $B$ is the transverse focusing strength, $U$ is the inter-vane voltage, and $\varphi_s$ is the synchronous phase.

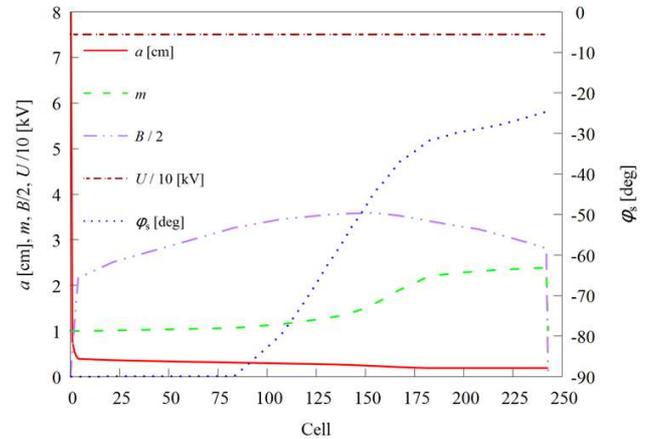

Figure 2: Main design parameters of the proton RFQ.

The beam dynamics simulation of the proton RFQ has been performed using the PARMTEQM code [11] with a 4D-Waterbag input distribution including $10^5$ macro-particles.

Fig. 3 shows the evolution of the important ratios used for the Hofmann charts, where the red curve stands for the tune ratio $\frac{\sigma_l}{\sigma_t}$, and the green and blue ones are for the transverse and longitudinal tune depressions, $\frac{\sigma_t}{\sigma_{0t}}$ and $\frac{\sigma_l}{\sigma_{0l}}$, respectively.

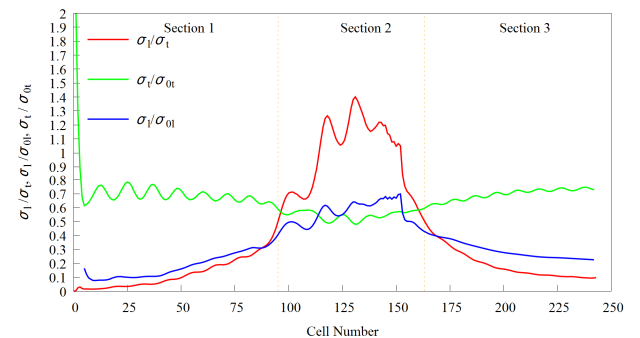

Figure 3: Evolution of tune ratio and tune depressions along the proton RFQ.

It can be seen that the tune ratio becomes larger than 0.5 at Cell 95 and becomes again smaller than 0.5 at Cell 163, which means the beam trajectories enter and leave the "safe rectangle" at Cell 95 and Cell 163, respectively. According to this, the proton RFQ can be divided into three sections: Section 1 (RFQ entrance to Cell 95), Section 2 (Cell 95 to Cell 163), and Section 3 (Cell 163 to RFQ exit).

In Fig. 4, the emittance ratio $\frac{\varepsilon_l}{\varepsilon_t}$ is plotted as a function of cell number. The $\frac{\varepsilon_l}{\varepsilon_t}$ curve has a "jump" around Cell 150, which is caused by less than 1% of particles outside of the separatrix. After they are lost, the $\frac{\varepsilon_l}{\varepsilon_t}$ curve comes back to the normal situation again.

Following the strategy for taking advantage of the emittance transfer, the emittance ratio at Cell 75 namely after the pre-bunching has been chosen as 1.3.

Afterwards, the emittance transfer occurs from the longitudinal phase plane to the transverse ones, so $\frac{\varepsilon_l}{\varepsilon_t}$ is being decreased. At the end of Section 1, it reaches 1.1.

In Section 2, the emittance transfer is low due to the "safe rectangle", so the $\frac{\varepsilon_l}{\varepsilon_t}$ value can be held around 1.1.

In the third section, the emittance transfer is getting stronger again, but slowly. The transfer direction is now from the transverse planes to the longitudinal one.

At the RFQ exit, both transverse and longitudinal emittances return to the levels at the beginning of the main bunching.

On the whole, the $\frac{\varepsilon_l}{\varepsilon_t}$ value is well inside the range from 0.9 to 1.4 for most positions along the proton RFQ.

In this way, the goal of using low emittance transfer to minimize emittance growth is reached.

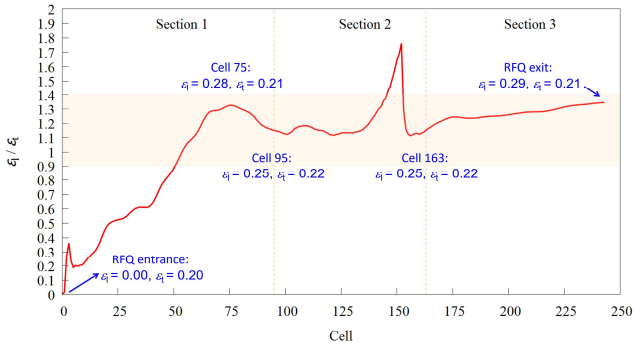

Figure 4: Emittance ratio $\frac{\varepsilon_l}{\varepsilon_t}$ as a function of cell number (all emittances in the figure are nomalized, rms values with a unit of π mm mrad).

## COMPARISON WITH THE EP METHOD

The main simulation results of the proton RFQ are summarized in Table 2. The RFQ length $L$ is about 3 m and the beam transmission efficiency is 99.1%. Both are very comparable to those of the J-PARC epRFQ.

As the J-PARC epRFQ uses a non-constant inter-vane voltage $U$, one needs to find an equivalent value for the comparison.

The shunt impedance of an RFQ, $R_p$, is defined as:

$$R_p = \frac{U^2 \times L}{P_c} \quad (2)$$

where $P_c$ is the RF power consumption. The nominal $P_c$ for the J-PARC epRFQ is 380 kW [12]. As $R_p$ is mainly inversely proportional to the radio frequency [13], one can assume that the J-PARC epRFQ has the same $R_p$ as its predecessor, the J-PARC RFQ III [14], which is also working at 324 MHz. For the J-PARC RFQ III, $P_c$ = 400 kW, $U$ = 81 kV, and $L$ = 3.623 m [14, 9]. Based on all these data, the calculated equivalent inter-vane voltage for the J-PARC epRFQ is 85.7 kV.

Using ~14% lower inter-vane voltage, the proton RFQ reaches smaller output emittance values in both transverse and longitudinal planes (see Table 2).

It can be seen in [9] that the transverse emittance is gradually but slowly increasing along the J-PARC epRFQ. Indeed, most of the beam trajectories in the J-PARC epRFQ have been successfully concentrated with the EP line as the focus. However, the beam trajectories cannot stay on the EP line exactly. As they are intensively oscillating around the EP line, the "killed" resonance peak will regrow from time to time. Although this peak is not very significant, the beam is going through it repeatedly so that the resonance can be accumulated.

Table 2: Design results of the proton RFQ

| Parameter | Proton RFQ | J-PARC epRFQ [9] |
|---|---|---|
| Input distribution | Waterbag | Waterbag |
| Inter-vane voltage [kV] | 75 | 61.3 – 143 |
| Input emittance $\varepsilon_{t,\,in,\,n,\,rms}$ [π mm mrad] | 0.20 | 0.20 |
| Onput emittance $\varepsilon_{t,\,out,\,n,\,rms}$ [π mm mrad] | 0.21 | 0.24 |
| Onput emittance $\varepsilon_{l,\,out,\,rms}$ [π MeV deg] | 0.10 | 0.11 |
| RFQ length [m] | 3.067 | 3.073 |
| Transmission [%] | 99.1 | 99.1 |

## CONCLUSION

A new design guideline is being proposed to minimize the space-charge-induced emittance growth by holding $\frac{\varepsilon_l}{\varepsilon_t}$ around one (in the range 0.9 – 1.4). It is not aiming to avoid the emittance transfer completely which is actually impossible in the real world, but suggesting to take advantage of low emittance transfer to realize designs with minimum emittance growth.

Furthermore, because the new guideline doesn't force the beam trajectories to stay on or closely around the EP line, it can change the beam dynamics parameters more quickly with more freedom. Therefore, it is promising to provide an efficient beam motion with low emittance growth for an RFQ accelerator.

## ACKNOWLEDGMENT

The author would like to sincerely thank Eugene Tanke for his kind help with the development of useful tools for data analyses.

# REFERENCES

[1] I. M. Kapchinskij and V. V. Vladimirskij, "Limitations of proton beam current in a strong focusing linear accelerator associated with the beam space charge", in *Proc. International Conference on High Energy Accelerators and Instrumentation*, CERN, Geneva, Sept. 1959, pp. 274-287.

[2] R. Chasman, "Numerical Calculations of the Effects of Space Charge on Six Dimensional Beam Dynamics in Proton Linear Accelerators", in *Proc. Proton Linear Accelerator Conference (LINAC'68),* New York, USA, May 1968, pp. 372-388.

[3] L. Smith, R. W. Chasman, K. R. Crandall, R. L. Gluckstern, T. Nishikawa, J. Haimson, P. M. Lapostolle, "Round Table Discussion of Space Charge and Related Effects", in *Proc. 6th Linear Accelerator Conference (LINAC'68)*, New York, USA, May 20-24, 1968, pp. 433-444.

[4] I. Hofmann, "Emittance Growth of Beams Close to the Space Charge Limit", in *Proc. 9th IEEE Particle Accelerator Conference*, Washington D.C., USA, Mar. 11-13 1981, pp. 2399-2401.

[5] I. Hofmann, I. Bozsik, "Computer Simulation of Longitudinal Transverse Space Charge Effects in Bunched Beams", in *Proc. 1981 Linac Accelerator Conference*, Santa Fe, New Mexico, USA, September 19-23, 1981, pp. 116–119.

[6] I. Hofmann, "Stability of anisotropic beams with space charge", Phys. Rev. E 57, 4 (1998) pp. 4713–4724.

[7] R. A. Jameson, "Equipartitioning in Linear Accelerators", in *Proc. 1981 Linac Accelerator Conference*, Santa Fe, New Mexico, USA, September 19-23, 1981, pp. 125–129.

[8] C. Zhang, H. Podlech, "Design Approach for a 325 MHz, 3 MeV, 70–100 mA Proton Radio-Frequency Quadrupole Accelerator with Low Emittance Transfer", Nucl. Instrum. Methods Phys. Res., Sect. A 947 (2019) 162756. `doi: 10.1016/j.nima.2019.162756`

[9] Y. Kondo, T. Morishita, R. A. Jameson, "Development of a radio frequency quadrupole linac implemented with the equipartitioning beam dynamics scheme", Phys. Rev. Accel. Beams 22 (2019) 120101.
`doi: 10.1103/PhysRevAccelBeams.22.120101`

[10] C. Zhang, A. Schempp, "Beam Dynamics Studies on a 200 mA Proton Radio Frequency Quadrupole Accelerator", Nucl. Instrum. Methods Phys. Res., Sect. A 586 (2008) 153-159. `doi: 10.1016/j.nima.2007.12.001`

[11] Manual of the LANL RFQ Design Codes, LANL Report No. LA-UR-96-1836 (revised June3, 2005).

[12] T. Morishita, Y. Kondo, "Electromagnetic design and tuning of the four-vane radio frequency quadrupole with nonuniform intervane voltage profile", Phys. Rev. Accel. Beams 23 (2020) 111003.
`doi: 10.1103/PhysRevAccelBeams.23.111003`

[13] C. Zhang, H. Podlech, "Efficient Focusing, Bunching, and Acceleration of High Current Heavy Ion Beams at Low Energy", Nucl. Instr and Meth. in Phys. Res. A 879 (2018) 19–24. `doi: 10.1016/j.nima.2017.09.059`

[14] T. Morishita, Y. Kondo, T. Hori, S. Yamazaki, K. Hasegawa, K. Hirano, H. Oguri, A. Takagi, T. Sugimura, F. Naito, "High-Power Test Results of the RFQ III in J-PARC Linac", in *Proc. 27th Linear Accelerator Conference (LINAC2014)*, Geneva, Switzerland, August 31-September 5, 2014, pp. 649-652.